\documentclass[aps,prb,twocolumn,10pt,superscriptaddress]{revtex4-2} 
\usepackage[utf8]{inputenc}
\usepackage{dsfont}
\usepackage{physics}
\usepackage{color}
\usepackage{hyperref}
\usepackage{stmaryrd}
\hypersetup{
    colorlinks=true,
    linkcolor=purple,
    citecolor=teal,     
    urlcolor=cyan,
    pdftitle={},
}
\usepackage{orcidlink}
\usepackage{soul}

\newcommand{\unit}[1]{\ensuremath{ \mathrm{#1}}}

\begin{document}

\title{Coherent errors in stabilizer codes caused by quasistatic phase damping}

\author{D\'{a}vid Pataki\,\orcidlink{0000-0002-8818-3723}}
\affiliation{Department of Theoretical Physics, Institute of Physics, Budapest University of Technology and Economics, M\H{u}egyetem rkp. 3., H-1111 Budapest, Hungary}

\author{\'{A}ron M\'{a}rton}
\affiliation{Department of Theoretical Physics, Institute of Physics, Budapest University of Technology and Economics, M\H{u}egyetem rkp. 3., H-1111 Budapest, Hungary}

\author{J\'{a}nos K. Asb\'{o}th\,\orcidlink{0000-0003-0962-7523}}
\affiliation{Department of Theoretical Physics, Institute of Physics, Budapest University of Technology and Economics, M\H{u}egyetem rkp. 3., H-1111 Budapest, Hungary}
\affiliation{HUN-REN Wigner Research Centre for Physics, H-1525 Budapest, P.O. Box 49., Hungary}

\author{Andr\'{a}s P\'{a}lyi}
\affiliation{Department of Theoretical Physics, Institute of Physics, Budapest University of Technology and Economics, M\H{u}egyetem rkp. 3., H-1111 Budapest, Hungary}
\affiliation{HUN-REN--BME Quantum Dynamics and Correlations Research Group, M\H{u}egyetem rkp. 3., H-1111 Budapest, Hungary}

\begin{abstract}
    Quantum error correction is a key challenge for the development of practical quantum computers, a direction in which significant experimental progress has been made in recent years.  In solid-state qubits, one of the leading information loss mechanisms is dephasing, usually modelled by phase flip errors. Here, we introduce quasistatic phase damping, a more subtle error model which describes the effect of Larmor frequency fluctuations due to $1/f$ noise. We show how this model is different from a simple phase flip error model, in terms of multi-cycle error correction. Considering the surface code, we provide numerical evidence for an error threshold, in the presence of quasistatic phase damping and readout errors. We discuss the implications of our results for spin qubits and superconducting qubits.
\end{abstract}

\maketitle

\tableofcontents

\section{Introduction} \label{introduction}

Quantum computing \cite{Nielsen_Chuang} promises efficient solutions to technologically relevant computational problems, but the realization of this great potential is hindered by errors in the physical implementation. 
Quantum error correction, with a thirty-year history from the fundamentals \cite{Shor_1995,Nielsen_Chuang} 
to the most recent experimental milestones \cite{PhysRevX.11.041058,Egan_2021,Linke_2017,Nigg_2014,Abobeih_2022,Bluvstein_2022,Bluvstein_2023},
offers a potential solution to this by the redundant encoding of a small number of logical qubits in a greater number of physical qubits.
The effect of Pauli errors during quantum error correction is well-understood due to the efficient simulability \cite{Aaronson_2004} and the existence of mappings to classical disordered spin models \cite{Dennis_2002,Wang_2003,Katzgraber_2009,Bombin_2012,Chubb_2021}. However, the exploration of more realistic error models is an ever-present challenge.

Recently, efficient large-scale numerical simulations of coherent errors became feasible, due to a newly developed technique \cite{Bravyi_2018}. This technique was originally used to investigate the effect of uniform coherent errors in the surface code, then follow-up works extended this to arbitrary planar-graph surface codes \cite{VennPRR2020} and the presence of phenomenological readout errors \cite{Marton2023coherenterrors}. These studies have numerically established the threshold behavior of the surface code in the presence of coherent errors, assuming uniform unwanted rotations on the data qubits.
 
In this work, we study the performance of Pauli stabilizer codes against \emph{quasistatic phase damping}. In this error model, physical qubits suffer coherent Z-rotations, with rotation angle randomly varying from qubit to qubit, but timewise constant throughout the entire quantum circuit; however, for each repetition of the circuit (shot) a new random value for each angle is chosen.
(i) We identify quantum computing architectures where this error model is relevant, and describes $1/f$ noise. 
(ii) For Pauli stabilizer codes \cite{gottesman1997stabilizer} we analyze how this error model is related to independent Pauli phase-flip errors. For a single cycle of error detection or error correction, we find the two error models are equivalent. For multiple syndrome measurement cycles this is no longer true, however, quasistatic phase damping is an effective -- correlated -- Pauli model on the logical level. 
It is known that for single qubit error models stabilizer measurements asymptotically decohere the noise on the logical level \cite{PhysRevLett.121.190501}. Our results show that for quasistatic phase damping this decoherence process is exact for all code sizes.
(iii) We utilize the Fermionic Linear Optics simulation framework based on the Majorana-fermionic representation of the surface code \cite{Bravyi_2018}, combined with phenomenological readout errors \cite{Marton2023coherenterrors}, to perform numerical experiments predicting the performance of the surface code in the presence of the quasistatic coherent error model.
(iv) We establish a surface code error-correction threshold at a physical error rate $p_\text{th} \approx 2.85 \%$, including the quasistatic coherent errors and the readout errors. 
(v) We generate the threshold phase diagram in the plane spanned by the strengths of the coherent error and the readout error. 
(vi) We discuss the implications of our results for spin qubits and superconducting qubits. 

The rest of this paper is structured as follows. In Sec.~\ref{sec-2-model}, we introduce the error model of quasistatic phase damping and pinpoint the relevant qubit platforms affected by this type of noise. In Sec.~\ref{sec-3-repetition-code}, we compare and contrast our error model with independent Pauli errors through a minimal example of the 2-qubit phase-flip repetition code. We generalize this comparison for Pauli stabilizer codes in Sec.~\ref{sec-4}. In Sec.~\ref{sec-5-surface-code} and Sec.~\ref{sec-6}, we present our results on the combined effect of quasistatic phase damping and readout errors on the surface code, and relate these results to semiconductor spin qubits and superconducting qubits.

\section{Quasistatic phase damping as a model of random coherent errors} \label{sec-2-model}

In this section, we introduce \emph{quasistatic phase damping}, the model of random coherent errors we study in the rest of this work.
We motivate this error model by its experimental relevance for state-of-the-art solid-state quantum computing platforms: quantum-dot-based semiconductor spin qubits and superconducting qubits.

The quasistatic phase damping error model consists of coherent rotations around the $z$ axis with a random angle for each data qubit. 
It is described by the unitary operator
\begin{equation}
\label{eq:quasistaticerror}
    \hat{U} = \prod_{j=1}^n e^{i\theta_j \hat{Z}_j}.
\end{equation}
where $n$ is the number of data qubits, and $\theta_j$ is the rotation angle of qubit $j$. 
We assume that in a single shot of a multi-cycle error detection or correction experiment, the random angles $\theta_j$ do not change for subsequent cycles.
However, in a multi-shot experiment, which is required to obtain a statistical evaluation of the error correction protocol, the angles $\theta_j$ change randomly from shot to shot.
These two properties are referenced by the adjective \emph{quasistatic}.
For concreteness, we assume that the angles $\theta_j$ are drawn randomly from a Gaussian distribution with zero mean and standard deviation $\sigma$, such that the random values for different qubits and different shots are statistically independent.
Note that this model can be regarded as an extension of Refs.~\cite{Bravyi_2018,VennPRR2020,Venn_2023,Marton2023coherenterrors}, where the effect of homogeneous coherent $Z$ rotations was studied in stabilizer codes.
We also note that the simulation method we use to treat this error model in this work generalizes in a straightforward way to more complicated angle distributions, including spatially and temporally correlated ones; that extension is beyond the scope of this work.

Let us now motivate this error model by pointing out its relation to solid-state qubits. 
Superconducting quantum computer prototypes hosting around 100 qubits are available, and small-scale quantum error detection and correction circuits based on Pauli stabilizer codes have been implemented \cite{Takita,Gong,Zhao,Andersen,Marques,Krinner_2022,GoogleQuantum_2023}. 
Qubit-level errors in these devices can be classified as SPAM errors, gate errors, and idle errors. 
Insight into the physical mechanism of idle errors is gained routinely via single-qubit Ramsey, Hahn echo, or more complex dynamical decoupling experiments. 

In certain cases, the inhomogeneous dephasing time $T_2^*$ of a superconducting qubit, measured by the Ramsey experiment, is exceeded by the qubit relaxation time $T_1$; 
furthermore, it is common to find that the Hahn echo improves qubit coherence, i.e., the corresponding decoherence time $T_{2,\text{echo}}$ exceeds $T_2^*$ \cite{NakamuraPRL2002,Yoshihara,Kakuyanagi}. The coexistence of these two features, i.e., (i) a dephasing-limited coherence time, and (ii) a functioning Hahn echo, is consistent with, and often interpreted as evidence for, classical electromagnetic noise that induces Larmor-frequency fluctuations.
Furthermore, it is often found that the corresponding noise spectrum depending on the frequency $f$ is proportional to $1/f$.

The quasistatic phase damping model \cite{Ithier,Paladino} we study here (see Eq.~\eqref{eq:quasistaticerror}) describes such Larmor frequency fluctuations of idling data qubits in a simplified manner. 
Since $1/f$ noise is dominated by the low-frequency range of the spectral density, it is reasonable to assume that the noisy component $\xi_j(t)$ of the Larmor frequency of qubit $j$ is constant throughout a single shot of a multi-cycle quantum error correction circuit, which typically takes a few microseconds. 
However, to obtain a statistical evaluation of the error correction performance, many shots should be taken, and the corresponding total data acquisition time window is sufficiently long that the noisy component $\xi_j$ explores many random values, represented by a Gaussian distribution in our model. 

A further assumption of our model is the statistical independence of the local random components of the Larmor frequencies, which is realistic if the Larmor frequency fluctuations are caused by local noise sources (e.g., gate electrodes, local flux bias lines, short-range charge fluctuators, etc.). We note that the numerical methods applied in this work are directly generalizable to variants of our noise model with more complex temporal and spatial correlations, which is in fact an interesting direction for future research.

The quasistatic phase damping model is relevant for semiconductor spin qubits as well, since spin qubit decoherence is often dominated by $1/f$ noise. \cite{Paladino,Burkard_2023,Struck_2020,Yoneda}
Note that in such devices, $1/f$ noise is attributed either to externally imposed electromagnetic fluctuations, e.g., gate noise, magnet noise, fluctuating charge traps, or to hyperfine interaction, i.e., magnetic noise exerted on the electron spin by the nuclear spins covered by the electron. 

In this work, we consider stabilizer codes as a way of countering quasistatic phase damping on idling qubits. 
As mentioned above, Hahn echo, and other dynamical decoupling schemes provide established alternative solutions for the same problem. 
An important and timely research direction is to compare these solutions, and to explore their combinations \cite{Krinner_2022,Zhao,GoogleQuantum_2023} to maximize the coherence of logical qubits.

\section{Example: 2-qubit repetition code} \label{sec-3-repetition-code}

\begin{figure*}[ht]
    \centering
    \includegraphics[width=0.75\textwidth]{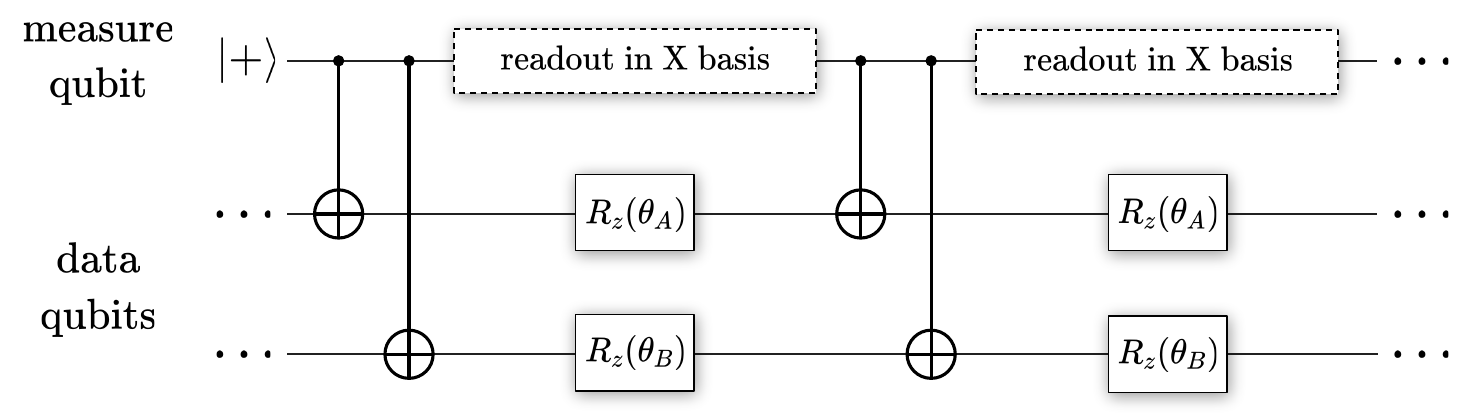}
    \caption{Schematic circuit representation of the $2$-qubit phase-flip repetition code, data qubits A and B suffering quasistatic phase damping. The figure shows elements of two subsequent cycles in a multi-cycle error detection procedure.
    The gates labelled as $R_z(\theta_{A,B})$ represent coherent errors.}
\label{fig:repetition-code-setup}
\end{figure*}

In this section, we study quasistatic phase damping using the example of a 2-qubit phase-flip code,  as depicted on Fig.~\ref{fig:repetition-code-setup}. 
We start with the analysis of a single cycle of syndrome measurement in Sec.~\ref{subsec-3A}, and show that in this case, quasistatic phase damping is equivalent to a simple uncorrelated phase-flip channel. In Sec.~\ref{subsec-3B}, explicit calculation of the density matrix reveals that this is not the case for two cycles of error detection. We quantify the difference between our model and the simple Pauli phase-flip model in Sec.~\ref{subsec-3C}. 
This simple example proves that quasistatic phase damping is a type of error that is distinct from phase-flip Pauli errors. This observation motivates the study of the stabiliser codes subject to quasistatic phase damping, which we do, focusing on the surface code, in Secs.~\ref{sec-5-surface-code} and \ref{sec-6}.

In the two-qubit phase-flip repetition code, errors are detected by repeatedly measuring a single \emph{parity-check} (or \emph{stabilizer}) operator $\hat{S}$. After each measurement, the state of the two qubits is altered by  projectors $\hat{\Pi}_s$ that depend on the measurement outcome $s \in \{+1,-1\}$ (also called even or odd parity, respectively),
\begin{align}
    \hat{\Pi}_s &= \dfrac{1}{2}(\hat{\mathds{1}} + s \hat{S});& 
    \hat S &= \hat{X}_A \hat{X}_B;& s\in \{+1,-1\}.
\end{align}
Here, $A$ and $B$ label the two data qubits.
The protocol aims to preserve the value of the logical $\hat{Z}$ operators of the code, which reads,  
\begin{align}
    \hat{Z}^L &= \hat{Z}_A \hat{Z}_B. 
\end{align}

\subsection{Single cycle of error detection} \label{subsec-3A}

A single cycle of the two-qubit phase-flip repetition code, as an error detection protocol, is capable of detecting a single Pauli Z error. The list of all possible 2-qubit Pauli Z error strings is brief: 
\begin{align}
    \hat{E}\in\{ \hat{\mathds{1}}, \hat{Z}_A, \hat{Z}_B, \hat{Z}_A \hat{Z}_B  \}.
    \label{eq:2qubit_single_error}
\end{align}

\subsubsection{Independent phase-flip errors}

We first consider independent phase-flip errors on the two physical qubits constituting the code. If the error probability on both qubits is  $p$, the Pauli error channel of the qubits reads,
\begin{align}
    \mathcal{E}_p (\hat{\rho}_0) &= \sum_{\hat{E}} P(\hat{E}) \hat{E} \hat{\rho}_0 \hat{E} = (1-p)^2 \hat{\rho}_0 + p(1-p) \hat{Z}_A \hat{\rho}_0 \hat{Z}_A \nonumber \\ 
    &\quad + (1-p)p \hat{Z}_B \hat{\rho}_0 \hat{Z}_B + p^2 \hat{Z}_A \hat{Z}_B \hat{\rho}_0 \hat{Z}_A \hat{Z}_B ,
\end{align}
for an arbitrary logical state of the two qubits $\hat{\rho}_0 = \ketbra{\psi_L} $, where $\ket{\psi_L}$ is an eigenstate of $\hat{S}$, 
\begin{equation}
    \hat{S}\ket{\psi_L} = \ket{\psi_L}.
\end{equation}
After the stabilizer measurement, for the trivial $s=+1$ outcome, the 
state of the two data qubits is projected to
\begin{equation}\label{eq:rep-2-rho1}
    \hat{\rho}_{1,p} (+1) = \hat{\Pi}_+ \mathcal{E}_p (\hat{\rho}_0) \hat{\Pi}_+ =  (1-p)^2 \hat{\rho}_0 + p^2 \hat{Z}^L \hat{\rho}_0 \hat{Z}^L.
\end{equation}

\subsubsection{Quasistatic phase damping}

We next consider independent coherent errors on the two qubits. The corresponding quantum channel reads,
\begin{equation}
    \mathcal{E}_\text{coh} (\hat{\rho}_0) = \hat{U} \hat{\rho}_0 \hat{U}^\dagger, \quad \text{with} \quad \hat{U} = e^{i \theta_A \hat{Z}_A} e^{i \theta_B \hat{Z}_B}.
\end{equation}
As mentioned in the previous section, we sample the rotation angles from  
Gaussian distributions with zero mean, and identical standard deviations $\sigma$, i.e., with probability density function $f(\theta_A,\theta_B)$, which reads, 
\begin{align}
f(\theta_A,\theta_B)&=f(\theta_A)f(\theta_B);&
f(\theta) &= \frac{1}{\sqrt{2\pi\sigma^2}} 
e^{ -\frac{\theta^2}{2\sigma^2}}, 
\end{align}
for $-\pi<\theta \le \pi$. 
Note that in this single-cycle setting, our quasistatic phase damping model is the same as phase damping, as described in Sec.~8.3.6 of \cite{Nielsen_Chuang}, independently on each qubit.

In this case, for the $s=+1$ measurement outcome, the density matrix of the two data qubits, averaged over the random angle distribution, reads
\begin{align}
    \hat{\rho}_{1,{\text{coh}}} (+1) &= 
    \int \mathrm{d} \theta_A \mathrm{d}\theta_B 
    f(\theta_A,\theta_B) 
    \hat{\Pi}_+ \mathcal{E}_\text{coh} (\hat{\rho}_0) 
    \hat{\Pi}_+ \nonumber \\
    \quad &\quad= 
    \langle \cos^2\theta \rangle^2 \hat{\rho}_0 
    + \langle\sin^2\theta \rangle^2 \hat{Z}^L \hat{\rho}_0 \hat{Z}^L,  \label{eq:rep-2-rho1-tilde}
\end{align}
where $\langle  .  \rangle$ denotes averaging over the random angle distribution:
\begin{align}
    \langle g(\theta) \rangle = \int_{-\pi}^{\pi} \mathrm{d}\theta g(\theta) f(\theta).
\end{align}

\subsubsection{Average effect of coherent errors can be represented as a random Pauli channel}

Comparing Eqs.~(\ref{eq:rep-2-rho1}) and (\ref{eq:rep-2-rho1-tilde}) we find that for a single cycle of error detection, choosing
\begin{equation}\label{eq:rep-2-single-cycle-p}
    p= \langle\sin^2\theta\rangle =\dfrac{1}{2}\left(1 - e^{-2 \sigma^2} \right),
\end{equation}
the two density matrices are identical,
\begin{equation}
    \hat{\rho}_{1,p}(+1) = \hat{\rho}_{1,\text{coh}} (+1) .
\end{equation}
This is also true for the syndrome $s=-1$. Hence, we conclude that in the case of a single cycle of error detection, the effect of uncorrelated random Pauli phase-flip errors is equivalent to the effect of phase damping.
This simple equivalence is broken in the multi-cycle case, as we show below.

\subsection{Two cycles of error detection} \label{subsec-3B}

We now consider two cycles of error detection. 
There are altogether 4 different measurement outcomes, \emph{syndromes}, in the two cycles, 
\begin{equation}
    \boldsymbol{s} \in \{(+1,+1), (-1,+1), (+1,-1), (-1,-1) \}.
\end{equation}
There are 
16 different cycle-resolved error scenarios $\boldsymbol{E}$; these error scenarios are the configurations of $Z$ error strings occurred in the different cycles: 
\begin{align}
    \boldsymbol{E} = (\hat{E}_1, \hat{E}_2), 
\end{align} 
with the two error strings $\hat{E}_1$ and $\hat{E}_2$ occurring in the 1st and 2nd cycles, respectively, are elements of the error set defined in Eq.~\eqref{eq:2qubit_single_error}. 
We denote the \emph{merged} error operators as 
\begin{align}
    \hat{E}(\boldsymbol{E}) = \hat{E}_1 \hat{E}_2. 
\end{align}
The set of the merged error operators is the same as the error set defined in Eq.~\eqref{eq:2qubit_single_error}.
The list of the 16 cycle-resolved error scenarios and corresponding merged error operators is given in Table~\ref{tab:rep-2-errors} in the Appendix.

\subsubsection{Independent phase-flip errors}

For independent phase-flip errors, after two cycles of syndrome measurements, the density matrix can be written in a Pauli-diagonal form
\begin{equation}
\label{eq:rep-2-rho2}
    \hat{\rho}_2 (\boldsymbol{s}) =  \sum_{\boldsymbol{E}\in \mathcal{D}_{\boldsymbol{s}}} P(\boldsymbol{E}) \hat{E}(\boldsymbol{E}) \hat{\rho}_0 \hat{E}(\boldsymbol{E}),  
\end{equation}
where $\mathcal{D}_{\boldsymbol{s}}$ denotes the set of cycle-resolved error scenarios consistent with the syndrome $\boldsymbol{s}$.
For example, for the trivial syndrome $(+1,+1)$, there are 4 compatible scenarios: 
$\boldsymbol{E} \in \{(\hat{\mathds{1}},\hat{\mathds{1}}), (\hat{Z}^L,\hat{\mathds{1}}), (\hat{\mathds{1}},\hat{Z}^L), (\hat{Z}^L,\hat{Z}^L) \} $, 
corresponding to 2 different merged error strings only, $\hat{E}(\hat{\mathds{1}},\hat{\mathds{1}}) = \hat{E}(\hat{Z}^L,\hat{Z}^L) = \hat{\mathds{1}}$ and $\hat{E}(\hat{Z}^L,\hat{\mathds{1}}) = \hat{E}(\hat{\mathds{1}},\hat{Z}^L) = \hat{Z}^L$.
(Table \ref{tab:rep-2-errors} lists all merged error operators along with the syndrome each of them implies.) 
For this syndrome, Eq.~\eqref{eq:rep-2-rho2} implies:
\begin{align}
    \hat{\rho}_2 (+1,+1) &= \left( P(\hat{\mathds{1}},\hat{\mathds{1}}) 
    + P(\hat{Z}^L,\hat{Z}^L) \right) \hat{\rho}_0 \nonumber\\
    &+ \left( P(\hat{Z}^L,\hat{\mathds{1}}) 
    + P(\hat{\mathds{1}},\hat{Z}^L) \right) 
    \hat{Z}^L\hat{\rho}_0 \hat{Z}^L.
\end{align}

We introduce the notation 
${c}_{\boldsymbol{s}}$ and ${d}_{\boldsymbol{s}}$
for the coefficients of various operators in the post-measurement state, 
\begin{equation}\label{eq:rep-2-2-cycles}
    \hat{\rho}_2 (\boldsymbol{s}) \!=\!
    \begin{cases}
    {c}_{\boldsymbol{s}} \hat{\rho}_0 + {d}_{\boldsymbol{s}} \hat{Z}^L 
    \hat{\rho}_0 \hat{Z}^L , & \text{for } s_2 =+1; \\
    {c}_{\boldsymbol{s}} \hat{Z}_A \hat{\rho}_0 \hat{Z}_A + 
    {d}_{\boldsymbol{s}} \hat{Z}_B \hat{\rho}_0 \hat{Z}_B , 
    & \text{for } s_2 =-1. \\  
    \end{cases}
\end{equation}
The coefficients ${c}_{\boldsymbol{s}}, {d}_{\boldsymbol{s}}$ are positive and fulfill
\begin{equation}
 \sum_{\boldsymbol{s}} \left( {c}_{\boldsymbol{s}} 
 + {d}_{\boldsymbol{s}} \right) = 1  ,
\end{equation}
and thus can be interpreted as probabilities. 
From Eq.~(\ref{eq:rep-2-rho2}), we obtain 
\begin{subequations} \label{eq:2qubit_2cycles_randompauli}
    \begin{align}
    {c}_{++} &= (1-p)^4 + p^4 , \\
    {d}_{++} &= {c}_{-+} = {d}_{-+} = 2 p^2 (1-p)^2 , \\
    {c}_{+-} &= {d}_{+-} = {c}_{--} = {d}_{--} = p (1-p)^3 + p^3 (1-p),
\end{align}
\end{subequations}
where we have compactified the notation, e.g., $c_{++} \equiv c_{(+1,+1)}$.

\subsubsection{Quasistatic phase damping}
\label{subsubsec:phasedamping}

Next, we consider quasistatic phase damping, with random rotation angles $\theta_A$ and $\theta_B$ for the two qubits taken from the same Gaussian distribution, and the same values used before both measurement cycles, as illustrated in 
Fig.~\ref{fig:repetition-code-setup}. 
Before performing the averaging for the error angles, the effect of the coherent errors cannot be written as a Pauli diagonal channel as in Eq.~\eqref{eq:rep-2-rho2}.
However, the angle distribution is symmetric with respect to zero. 
Therefore, after averaging over the angle distribution according to the quasistatic phase damping model, we do obtain a Pauli diagonal channel, which can be written in the form of Eq.~(\ref{eq:rep-2-2-cycles}), although with different $\Tilde{c}_{\boldsymbol{s}}$ and $\Tilde{d}_{\boldsymbol{s}}$ coefficients - we use $\Tilde{}$ to specify the case of quasistatic phase damping:
\begin{subequations}
\begin{align}
    \Tilde{c}_{++} &= \frac{1}{16} \left(e^{-16 \sigma^2}+2 e^{-8 \sigma^2}+8 e^{-4 \sigma^2}+5\right) , \\
    \Tilde{d}_{++} &= \Tilde{d}_{-+} = \frac{1}{16} \left(1-e^{-8 \sigma^2}\right)^2 , \\
    \Tilde{c}_{-+} &= \frac{1}{16} \left(e^{-16 \sigma^2}+2 e^{-8 \sigma^2}-8 e^{-4 \sigma^2}+5\right) , \\
    \Tilde{c}_{+-} &= \Tilde{d}_{+-} = \Tilde{c}_{--} = \Tilde{d}_{--} = \frac{1}{16} \left(1-e^{-16 \sigma^2}\right) . 
\end{align}
    \label{eq:2qubit_2cycles_coherent}
\end{subequations}

From this result, we conclude that in this two-cyle case, the effect of quasistatic phase damping cannot be described as random independent phase-flip errors with the same error rate $p$ on both qubits. 
This is apparent by a comparison of Eqs.~\eqref{eq:2qubit_2cycles_randompauli} and \eqref{eq:2qubit_2cycles_coherent}: 
for independent homogeneous phase-flip errors, there are only 3 different ${c}_{\boldsymbol{s}}$, ${d}_{\boldsymbol{s}}$ coefficients, while for quasistatic phase damping, there are 4 different coefficients.

\subsubsection{Best approximation of quasistatic phase damping with independent homogeneous phase-flip errors} 
\label{subsec-3C}

As discussed just above, with two subsequent cycles of error detection in the two-qubit phase-flip code, the effect of quasistatic phase damping cannot be described as random independent phase-flip errors with the same error parameter $p$ on both qubits. 
In this section, we try to find the best independent homogeneous phase-flip approximation of the quasistatic phase damping channel.

We will quantify the quality of the approximation by the total variational distance (TVD).  
This distance is defined as 
\begin{align}
    \delta(p, \sigma) = \max_{\boldsymbol{s}} \left( \max \left\{ \left| 
    {c}_{\boldsymbol{s}} - \tilde{c}_{\boldsymbol{s}}\right|, \left| 
    {d}_{\boldsymbol{s}} - \tilde{d}_{\boldsymbol{s}}\right| \right\} \right),
\end{align}
where the right-hand-side depends on the single parameter of the random Pauli channel, $p$, through the coefficients $c_{\boldsymbol{s}}$ and $d_{\boldsymbol{s}}$, and on the single parameter of the quasistatic phase damping channel, the standard deviation of the coherent error distribution $\sigma$, through the coefficients $\tilde{c}_{\boldsymbol{s}}$ and $\tilde{d}_{\boldsymbol{s}}$. 
The best Pauli approximation (i.e., the best approximation with independent homogeneous phase-flip) of a quasistatic phase damping channel with  $\sigma$ has the parameter $p_\text{best}$, such that 
\begin{equation}
\label{eq:def_p_best}
    \delta(p_\text{best}(\sigma), \sigma) \leq \delta(p, \sigma) \quad \text{for all } p.
\end{equation}

\begin{figure}
    \centering
    \includegraphics[width=0.45\textwidth]{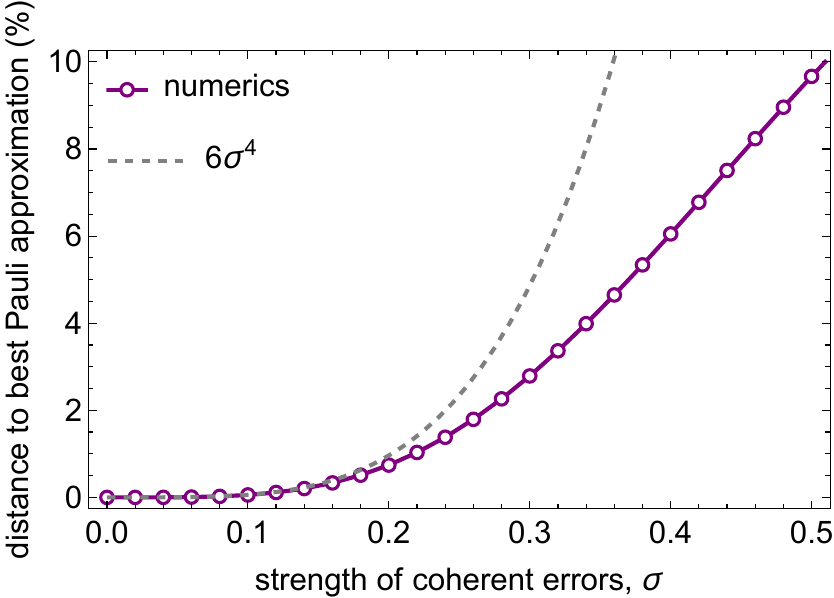}
    \caption{Distance to best Pauli approximation as a function of error parameter $\sigma$ for the $2$-qubit phase-flip repetition code.}   
\label{fig:repetition-code-TVDmin}
\end{figure}

We computed the best Pauli approximation numerically, and show the corresponding total variational distance in Fig.~\ref{fig:repetition-code-TVDmin}. 
For small $\sigma$, we find good agreement with the power series expansion,  
\begin{align} 
\label{eq:pbest-approx}
    \sigma\ll 1: \quad p_\text{best}(\sigma) &\approx \sigma^2;& 
    \delta(p_\text{best}, \sigma) &\approx 6 \sigma^4,
\end{align}
with less than a few percent distance between our noise model and the optimal single-parameter Pauli error model (see Fig.~\ref{fig:repetition-code-TVDmin}). Increasing $\sigma$, the distance to the best Pauli approximation is also increased, for $\sigma = 0.5$ it is around $10\%$. 

\section{Quasistatic phase damping in stabilizer codes} \label{sec-4}

In this section, we generalize the results above, i.e., compare the quasistatic phase damping (quasistatic random coherent error) model with the uncorrelated phase-flip model, for Pauli stabilizer codes in general. In Sec.~\ref{subsec-4A}, we prove that the two models are equivalent in the case of a single cycle of error detection or error correction. 
In Sec.~\ref{subsec-4B}, we show that on the logical level, quasistatic phase damping is equivalent to a Pauli channel, however, this is not the one resulting from the simple uncorrelated phase-flip error model in general (see previous section as an example).
Motivated by this difference of the error models, we will study the effect of quasistatic phase damping on the surface code in Secs.~\ref{sec-5-surface-code} and \ref{sec-6}.

\subsection{Single cycle of error detection and error correction} \label{subsec-4A}

Now let us consider any $\llbracket n,k,d \rrbracket$ Pauli stabilizer code, encoding $k$ logical qubits into $n$ physical qubits with code distance $d$, whose stabilizers are tensor products of Pauli matrices and they commute with each other \cite{gottesman1997stabilizer}. 
The stabilizer generators (independent parity checks) of the code 
are denoted by $\hat{S}_f$, with $f=1, \ldots, N$, with $N=n-k$. 
Upon measuring the stabilizer generators, we obtain the syndrome
\begin{equation}
    s = (s_1, s_2, \dots s_N ) \in \{+1,-1\}^N, 
\end{equation}
and the quantum state of the qubits is projected to the code space with the projector
\begin{equation}
    \hat{\Pi}_s = \prod_{f=1}^N \dfrac{1}{2}\left( 1 + s_f \hat{S}_f \right).
\end{equation}

We aim to investigate the form of the quasistatic coherent error channel combined with error detection. To this end, we have to perform an averaging over the random angle distribution for the post-measurement state. We will denote this average by
\begin{equation}
    \langle \dots \rangle_{\boldsymbol{\theta}}   \equiv \int \mathrm{d}\boldsymbol{\theta} f(\boldsymbol{\theta}) \dots ,
\end{equation}
where $f(\boldsymbol{\theta}) = f(\theta_A)f(\theta_B)\cdots f(\theta_n)$ is the probability density function of the independent random rotation angles. 
The following derivation is also true for any (discrete or continuous) probability distribution that is symmetric and has zero mean value, i.e., 
\begin{align}
f(\theta) &= -f(\theta), \quad \text{with } -\pi < \theta \le \pi.    
\end{align} 
Note that in this case, the average of any odd function of $\theta_j$ will vanish.

For a single cycle of error detection, the effect of quasistatic phase damping is expressed as:
\begin{align}
\label{eq:rho1stabilizer}
    \hat{\rho}_1 (s) &=  \big\langle\hat{\Pi}_s \hat{U} \hat{\rho}_0 \;\hat{U}^\dagger \hat{\Pi}_s \big\rangle_{\boldsymbol{\theta}} \nonumber \\
    &= \sum_{\hat{E}, \hat{E}^\prime} \Big\langle  A_{\boldsymbol{\theta}}(\hat{E})  A^\ast_{\boldsymbol{\theta}} (\hat{E}^\prime) \Big\rangle_{\boldsymbol{\theta}} \hat{\Pi}_s \hat{E} \hat{\rho}_0 \hat{E}^\prime \hat{\Pi}_s .
\end{align}
Here, $\hat{\rho}_0 = \ketbra{\psi_L}$ is the density matrix of an arbitrary initial logical state, i.e., with $\ket{\psi_L}$ being a common eigenstate of all stabilizer operators with +1 eigenvalue.
Furthermore, 
the $2^{n}$ error strings corresponding to $\hat{Z}$ phase-flip operations on either of the qubits are denoted as $ \hat{E}$,
and we have rewritten the product of coherent single qubit rotations as
\begin{equation}
    \hat{U} = \sum_{\hat{E}} A_{\boldsymbol{\theta}}(\hat{E}) \hat{E} ,
\end{equation}
by introducing the amplitudes
\begin{equation}\label{eq:P-tilde-theta}
    A_{\boldsymbol{\theta}}(\hat{E}) = \prod_{j=1}^n (\cos\theta_j)^{1-n_{\hat{E}} (j)} (i \sin\theta_j)^{n_{\hat{E}} (j)} .
\end{equation}
Here $n_{\hat{E}} (j) =0$ or $1$ denotes the number of Pauli Z operators acting on qubit $j$ in error string $\hat{E}$. 

Now, we show that Eq.~\eqref{eq:rho1stabilizer} can be simplified.
First, averaging over the random angles in Eq.~\eqref{eq:rho1stabilizer} yields:
\begin{equation}
\label{eq:AAaverage}
    \Big\langle  A_{\boldsymbol{\theta}}(\hat{E})  A^\ast_{\boldsymbol{\theta}} (\hat{E}^\prime) \Big\rangle_{\boldsymbol{\theta}} = \delta_{\hat{E},\hat{E}^\prime}  \Big\langle | A_{\boldsymbol{\theta}}(\hat{E}) |^2 \Big\rangle_{\boldsymbol{\theta}} \equiv P(\hat{E}).
\end{equation}
Second, we note that since the stabilizers are products of Pauli operators, they either commute or anti-commute with the error strings. Therefore applying the error string $\hat{E}$ and then measuring the syndrome $s$ compatible with that ($\hat{E} \in\mathcal{D}_s$) has the same effect as first projecting the collective quantum state of the qubits to the trivial syndrome (all parity checks are +1) and then applying the error string; i.e., it holds that 
\begin{equation}
\label{eq:stabilizertrick}
    \hat{\Pi}_{s} \hat{E} = \hat{E} \hat{\Pi}_{0} ,\quad\text{if}\quad \hat{E}\in\mathcal{D}_s.
\end{equation}

Hence, by exploiting Eqs.~\eqref{eq:AAaverage} and \eqref{eq:stabilizertrick} in Eq.~\eqref{eq:rho1stabilizer}, we obtain a simplified formula for the error channel,
\begin{equation}\label{eq:result-single-cycle}
    \hat{\rho}_1 (s) = \sum_{\hat{E} \in\mathcal{D}_s} P(\hat{E}) \hat{E} \hat{\rho}_0 \hat{E} ,
\end{equation}
where the $P(\hat{E})$ probabilities follow from Eq.~(\ref{eq:P-tilde-theta}) as
\begin{equation}
    P(\hat{E}) = \prod_{j=1}^n  \langle\cos^2\theta_j\rangle_{\boldsymbol{\theta}}^{1-n_{\hat{E}} (j)} \langle\sin^2\theta_j\rangle_{\boldsymbol{\theta}}^{n_{\hat{E}} (j)} .
\end{equation}
Since the $\hat{E}$ error strings are products of Pauli Z and identity matrices, the result, Eq.~(\ref{eq:result-single-cycle}), is a simple phase flip channel with equal and independent error probabilities on each qubit, $p = \langle\sin^2\theta_j\rangle_{\boldsymbol{\theta}}$, for each syndrome $s$. Consequently, for a single cycle of experiment performed with any Pauli stabilizer code, quasistatic phase damping is equivalent to uncorrelated phase-flip errors on average. Indeed that is what we expected, since for a single physical qubit, phase damping is the same as the phase flip channel \cite{Nielsen_Chuang}.

A result analogous to Eq.~\eqref{eq:result-single-cycle} holds for a single cycle of error \emph{correction} as well since the correction operation $\hat{C}_s$ depends only on the syndrome $s$.
Hence, the state of the code after error correction has the following form: 
\begin{equation}\label{}
    \hat{C}_s \hat{\rho}_1 (s)\hat{C}_s = \sum_{\hat{E} \in\mathcal{D}_s} P(\hat{E}) \hat{C}_s \hat{E} \hat{\rho}_0 \hat{E} \hat{C}_s .
\end{equation}

\subsection{Multiple cycles} \label{subsec-4B}

We next consider multiple, $t$ cycles of error detection: we apply the independent coherent rotations of the qubits with angles $\theta_1, \ldots, \theta_n$, and the measurements repeatedly. We show that in this case, the average effect of quasistatic phase damping can be represented only with a more complicated Pauli channel. It is important to note that the angles $\theta_j$ do not change from cycle to cycle: if they did, this error model would -- after averaging over the random angles -- be equivalent to independent phase-flip errors.

Altogether, there are $2^{t n}$ cycle-resolved error scenarios $\boldsymbol{E}$, corresponding to different combinations of  $\hat{Z}$ phase-flip operations on either of the qubits during either of the cycles, and $2^n$ different merged Pauli-Z error strings $\hat{E}(\boldsymbol{E})$,  
\begin{align}
\boldsymbol{E} &= \left( \hat{E}_1, \hat{E}_2, \dots  \hat{E}_t\right);&
\hat{E}(\boldsymbol{E})&= \prod_{r=1}^t \hat{E}_r.
\end{align}
With this notation, we express the syndrome-dependent state of the code after $t$ error detection cycles as follows:
\begin{align} \label{eq:rho-d-general}
    \hat{\rho}_t & (\boldsymbol{s}) = 
    \Big\langle \hat{Q}(\boldsymbol{\theta}) \hat{\rho}_0 
    \hat{Q}(\boldsymbol{\theta})^\dagger \Big\rangle_{\boldsymbol{\theta}}   
    \nonumber \\
    &= \sum_{\boldsymbol{E}\in \mathcal{D}_{\boldsymbol{s}}}  \sum_{\boldsymbol{E^\prime}\in \mathcal{D}_{\boldsymbol{s}}}  \Big\langle A_{\boldsymbol{\theta}}(\boldsymbol{E})  A^\ast_{\boldsymbol{\theta}} (\boldsymbol{E^\prime}) \Big\rangle_{\boldsymbol{\theta}}  \hat{E}(\boldsymbol{E}) \hat{\rho}_0 \hat{E} (\boldsymbol{E^\prime})  \nonumber\\
    &=  \sum_{\boldsymbol{E}\in \mathcal{D}_{\boldsymbol{s}}} \Tilde{P}(\boldsymbol{E}) \hat{E}(\boldsymbol{E}) \hat{\rho}_0 \hat{E}(\boldsymbol{E}),
\end{align}
where $\hat{Q}(\boldsymbol{\theta}) = 
    \hat{\Pi}_{s_t} \hat{U} 
    \hat{\Pi}_{s_{t-1}} \hat{U} \dots \hat{\Pi}_{s_1} \hat{U}$,     
and we introduced the notation
\begin{equation}
    A_{\boldsymbol{\theta}}(\boldsymbol{E}) = \prod_{r=1}^t A_{\boldsymbol{\theta}}(\hat{E}_r). 
\end{equation}

The density matrix in Eq.~\eqref{eq:rho-d-general} is again Pauli-diagonal, but the newly defined quantities
\begin{align} \label{eq:tilde-P-E}
    \Tilde{P}(\boldsymbol{E} \in \mathcal{D}_{\boldsymbol{s}}) &= \sum_{\boldsymbol{E^\prime}\in \mathcal{D}_{\boldsymbol{s}}} \Big\langle A_{\boldsymbol{\theta}}(\boldsymbol{E})  A^\ast_{\boldsymbol{\theta}} (\boldsymbol{E^\prime}) \Big\rangle_{\boldsymbol{\theta}} \\
    &= \sum_{\substack{\boldsymbol{E^\prime}\in \mathcal{D}_{\boldsymbol{s}} \\ \hat{E}(\boldsymbol{E}) = \hat{E}(\boldsymbol{E^\prime})}} \Big\langle A_{\boldsymbol{\theta}}(\boldsymbol{E})  A^\ast_{\boldsymbol{\theta}} (\boldsymbol{E^\prime}) \Big\rangle_{\boldsymbol{\theta}} ,
\end{align}
obtained by regrouping the terms in the double sum, are not probabilities: they add up to 1,
\begin{equation}
    \sum_{\boldsymbol{E}} \Tilde{P}(\boldsymbol{E}) =1,
\end{equation}
but they can be negative as well (see Table~\ref{tab:rep-2-errors} in the Appendix), depending on which power of the imaginary unit appears in the corresponding terms. Real values are guaranteed by the fact that the terms containing odd powers of $(i \sin\theta_j)$ will vanish after averaging.
In Eq.~\eqref{eq:tilde-P-E}, we only get a non-zero contribution when $\hat{E}(\boldsymbol{E}) = \hat{E}(\boldsymbol{E^\prime})$.

Upon performing the last sum in Eq.~\eqref{eq:rho-d-general}, we obtain the following result:
\begin{equation}
\label{eq:rhodlogical}
\hat{\rho}_t (\boldsymbol{s}) = 
\sum_{\boldsymbol{\alpha}} 
\Tilde{\mathcal{P}}_{\boldsymbol{\alpha}}(\boldsymbol{s}) \hat{Z}_{\boldsymbol{\alpha}} \hat{E}_{\boldsymbol{s}} 
\hat{\rho}_0 
\hat{E}_{\boldsymbol{s}}
\hat{Z}_{\boldsymbol{\alpha}},
\end{equation}
where 
\begin{equation}
\hat{Z}_{\boldsymbol{\alpha}} = (\hat{Z}^L_1)^{\alpha_1}(\hat{Z}^L_2)^{\alpha_2} \ldots (\hat{Z}^L_k)^{\alpha_k};
\end{equation}
\begin{equation}
    \Tilde{\mathcal{P}}_{\boldsymbol{\alpha}}(\boldsymbol{s}) = \sum_{\substack{\boldsymbol{E}\in \mathcal{D}_{\boldsymbol{s}} \\ \hat{E}(\boldsymbol{E}) = \hat{Z}_{\boldsymbol{\alpha}}\hat{E}_{\boldsymbol{s}}}}
    \Tilde{P}(\boldsymbol{E}).
\end{equation}
Furthermore, we defined 
$\boldsymbol{\alpha} = (\alpha_1, \ldots, \alpha_k) \in \{0, 1\}^{k}$, the operator $\hat{Z}^L_j$ is defined as the logical Z operator acting on the $j$-th logical qubit, and  $\hat{E}_{\boldsymbol{s}}$ is an arbitrary merged error string consistent with the syndrome $s$.
The $\Tilde{\mathcal{P}}_{\boldsymbol{\alpha}}(\boldsymbol{s})$ quantities are syndrome-dependent probabilities, meaning that they are positive (cf.~Eq.~(8.152) of \cite{Nielsen_Chuang}), and satisfy 
\begin{equation}
    \sum_{\boldsymbol{\alpha},\boldsymbol{s}} \Tilde{\mathcal{P}}_{\boldsymbol{\alpha}}(\boldsymbol{s}) = 1.
\end{equation}
These probabilities are the generalization of the single-logical-qubit coefficients $\Tilde{c}_{\boldsymbol{s}}$ and $\Tilde{d}_{\boldsymbol{s}}$ of Eq.~\eqref{eq:2qubit_2cycles_coherent}, e.g. $\Tilde{c}_{++} = \Tilde{\mathcal{P}}_{0}(++)$, $\Tilde{d}_{++} = \Tilde{\mathcal{P}}_{1}(++)$. As discussed in Sec.~\ref{subsec-3B}, these probabilities cannot be derived from independent single-qubit Z error probabilities.

We have seen that Eq.~\eqref{eq:rho-d-general} is not a well-defined Pauli channel on the physical level, because the $\Tilde{P}(\boldsymbol{E})$ weights can be negative, thus these are not probabilities.
However, we can rephrase Eq.~\eqref{eq:rho-d-general} by a new set of amplitudes $\Tilde{P}'(\boldsymbol{E})$ that are positive, and hence define a proper Pauli channel on the physical level as well.
This new set of amplitudes can be defined by redistributing the amplitudes within each class of cycle-resolved error scenarios ($\boldsymbol{E}$), where the classes are indexed by 
$\boldsymbol{s}$ and $\boldsymbol{\alpha}$, and are defined via $\boldsymbol{E}\in \mathcal{D}_{\boldsymbol{s}}$, $\hat{E}(\boldsymbol{E})=\hat{Z}_{\boldsymbol{\alpha}}\hat{E}_{\boldsymbol{s}}$. 
In words, two cycle-resolved error scenarios belong to the same class, if they imply the same syndrome and the same merged error string. 
By performing such a redistribution $\Tilde{P} \to \Tilde{P}'$ of amplitudes, the error channel in Eq.~\eqref{eq:rhodlogical} does not change: the $\Tilde{\mathcal{P}}_{\boldsymbol{\alpha}}(\boldsymbol{s})$ probabilities remain the same. Note that this redistribution is not unique; one way to do it is to choose one representative cycle-resolved error scenario $\boldsymbol{E}'_{\boldsymbol{s},\boldsymbol{\alpha}}$ from each error class, and define new amplitudes, which are in fact non-negative, as:
\begin{equation}
\Tilde{P}'(\boldsymbol{E}'_{\boldsymbol{s},\boldsymbol{\alpha}}) = \Tilde{\mathcal{P}}_{\boldsymbol{\alpha}}(\boldsymbol{s}),
\end{equation}
and $\Tilde{P}'(\boldsymbol{E}) = 0$ for all $\boldsymbol{E}$ that is not in the set of $\boldsymbol{E}'_{\boldsymbol{s},\boldsymbol{\alpha}}$ representatives.
This shows that our error model can be described as a Pauli model on the physical level. However, this Pauli model is highly correlated. It is an open question whether it is possible to construct an error model with local (in both space and time) correlations by redefining these weights differently. This question is beyond the scope of this work.

We conclude that for multiple cycles, the quasistatic phase damping channel is equivalent to an incoherent Pauli channel as written in Eq.~\eqref{eq:rho-d-general}, as well as an incoherent Pauli channel on the logical level, as written Eq.~\eqref{eq:rhodlogical}. 
However, the Pauli channel in Eq.~\eqref{eq:rho-d-general} does not describe, in general, independent homogeneous single-qubit phase flips, as we illustrated by the counterexample in Sec.~\ref{subsubsec:phasedamping}. 

\section{Surface code error threshold} \label{sec-5-surface-code}

So far, we have shown that for stabiliser codes applied for multi-cycle error correction, the effect of quasistatic phase damping is distinct from that of phase-flip Pauli errors. This motivates studies of the noise resilience in this new setting. In this section, we do this by focusing on the surface code. In particular, we numerically demonstrate that the surface code in the presence of quasistatic phase damping and readout errors exhibits a threshold behavior.
To this end, we apply the Fermionic Linear Optics (FLO) simulation technique \cite{Bravyi_2018,VennPRR2020,Marton2023coherenterrors}. In Sec.~\ref{subsec-5A} we briefly summarize the basics of the surface code and the effect of coherent errors. In Sec.~\ref{subsec-5B} we discuss error correction in the presence of readout errors. 
Finally, we present our numerical results in Sec.~\ref{subsec-5D}, showing that the threshold for quasistatic phase damping is, up to our numerical precision, the same as the threshold for Pauli phase flips, whereas the logical error rate at threshold is smaller in the the former case (7\%) than in the latter case (8.5\%).

\begin{figure}
    \centering
    \includegraphics[width=0.48\textwidth]{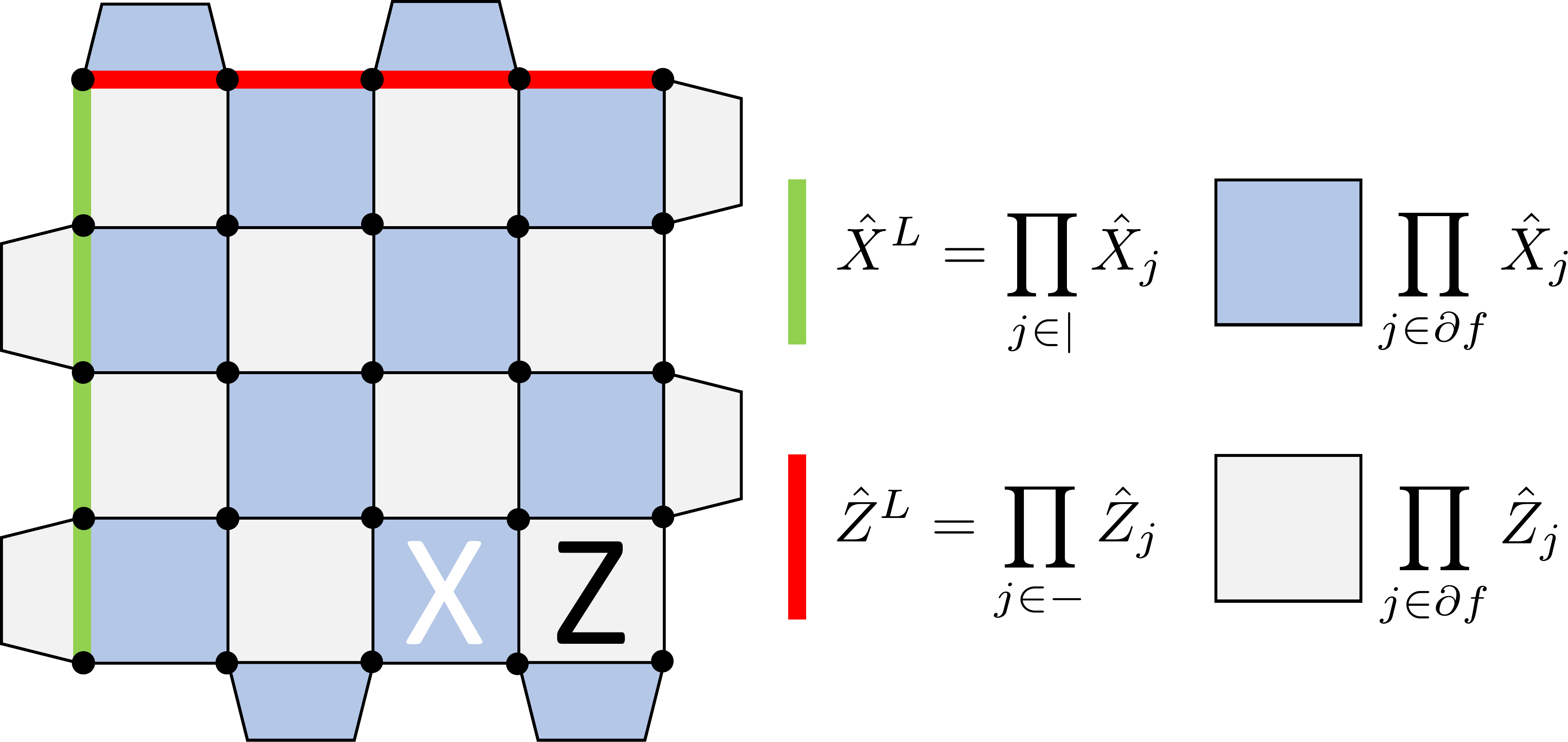}
    \caption{Layout of a distance 5 surface code patch. Blue (light gray) plaquettes represent X-type (Z-type) stabilizer operators. Green (red) string denotes a logical X (Z) operator.}
\label{fig:surface-code-setup}
\end{figure}

\subsection{Coherent errors on the physical qubits lead to coherent error on the logical qubit} \label{subsec-5A}

The surface code encodes a logical qubit in the common $+1$ eigenspace of the X-type and Z-type parity check operators (stabilizers), as shown in Fig.~\ref{fig:surface-code-setup}. Since we consider coherent $Z$ errors, only the $X$-type parity checks, 
$\hat{S}_f = \prod_{j\in\partial f} \hat{X}_j$,
can return a $-1$ value. 
As a result of a single cycle of measurements, which we assume to be free of readout error for now, a syndrome $s$ is obtained. 
If this syndrome $s$ is non-trivial, then a non-identity correction operation $\hat{C}_{s}$ is applied as an attempt to bring the state back into the code space. In this case, $\hat{C}_{s}$ involves a collection of Pauli Z phase flips. 
The specific correction operation for a given syndrome is the solution of the decoding problem, which we discuss briefly in Sec.~\ref{subsec-5B}.

Coherent errors on the physical qubits, upon error correction, lead to a coherent error of the same type on the logical level \cite{Marton2023coherenterrors}. 
Starting from an arbitrary logical state $\ket{\psi_L}$, the quantum state of $n=d^2$ data qubits after measuring syndrome $s$, and applying the corresponding correction, can be written as \cite{Bravyi_2018}
\begin{equation}
    \ket{\Phi_{s}} = \dfrac{1}{\sqrt{P(s)}}\hat{C}_{s}\hat{\Pi}_{s}\hat{U}\ket{\psi_L} = e^{i\theta_L \hat{Z}^L} \ket{\psi_L} ,
\end{equation}
where $P(s)$ is the probability of syndrome $s$, and the logical rotation angle $\theta_L$ depends on the syndrome, the decoder, and the physical rotation angles, $\theta_L = \theta_L (s,\boldsymbol{\theta})$, but \emph{not} on the initial state $\ket{\psi_L}$.

For a given syndrome $s$, we quantify the logical error by the maximum infidelity of the final state $\ket{\Phi_{s}}$ with the initial state $\ket{\psi_L}$, maximized over all initial states. Logical rotations around the $z$ axis are the most harmful for the eigenstates of the $\hat{X}^L$ operator, thus we obtain the logical error rate as
\begin{equation}
   p_L (s,\boldsymbol{\theta}) \equiv \max_{\ket{\psi_L}} \left( 1 - |\bra{\Psi_L}\ket{\Phi_{s}}|^2 \right) = \sin^2\theta_L(s,\boldsymbol{\theta}),
\end{equation}
with this maximum attained for $\ket{\Psi_L} = \ket{+_L}$, the common eigenstate of $\hat{X}^L$ as well as of all the stabilizer generators $\hat{S}_f$ with eigenvalue +1.

To characterize the strength of the quasistatic phase damping errors, we will use either the standard deviation $\sigma$ of the single-qubit random coherent rotations, or the \emph{physical error rate} $p$ which is defined as for the repetition code, Eq.~(\ref{eq:rep-2-single-cycle-p}).
This definition is motivated by the fact that for single-cycle error detection and error correction, it provides an exact correspondence with the uncorrelated phase-flip channel.

\subsection{Error correction with readout errors} \label{subsec-5B}

We use a phenomenological noise model for readout errors where the syndrome measurements are perfect, but the outcomes are unreliably recorded, with readout error probability $q$,
\begin{equation}
    P(-1\to +1) = P(+1\to -1) = q
\end{equation}
and $P(-1\to -1) = P(+1\to +1) = 1-q $. 
To solve the decoding problem in the presence
of readout errors, we consider $d$ consecutive cycles of syndrome measurements. As a result, we get the noisy multi-cycle syndrome
\begin{equation}
    \boldsymbol{s}= (s_1, s_2, \dots s_d ) \to  \boldsymbol{s}^\prime = (s_1^\prime, s_2^\prime, \dots s_d^\prime )   .
\end{equation}
The state of the code after measuring the  noisy multi-cycle syndrome $\boldsymbol{s}^\prime$, decoding it, and applying the corresponding correction operator, reads:
\begin{equation}    \ket{\Phi_{\boldsymbol{s},\boldsymbol{s^\prime}}} = \dfrac{1}{\sqrt{P(\boldsymbol{s})}}\hat{C}_{\boldsymbol{s^\prime}}\hat{\Pi}_{s_d}\hat{U}\dots\hat{\Pi}_{s_1}\hat{U}\ket{\psi_L} = e^{i\theta_L \hat{Z}^L} \ket{\psi_L} ,
\end{equation}
where the logical rotation angle $\theta_L$ now depends on the noisy multi-cycle syndrome as well, $\theta_L = \theta_L (\boldsymbol{s},\boldsymbol{s^\prime},\boldsymbol{\theta})$.

To correct errors based on the noisy multi-cycle syndrome that is contaminated by readout error,  we use the 3-dimensional version of the minimum weight perfect matching (MWPM) decoder \cite{fowler2014minimum}, as implemented in PyMatching \cite{higgott2023sparse}. 
The decoder sees the measurement outcomes as a 3-dimensional grid that has $(d^2-1)/2\times (d^2-1)/2$ `space' coordinates, characterizing the positions of the measured X stabilizers, and $d$ `time' coordinates, numbering the measurement cycles.
Vertices, where the measured parity check differs from the value measured in the previous cycle, are marked, and the decoder needs to identify the set of edges with the smallest weight that connects the marked vertices to the left/right boundaries or pairs them up. The correction operation is inferred from the resulting set of edges. `Spacelike' and `timelike' edges on the grid correspond to readout errors and physical errors. These carry different weights
\begin{equation}
    w_s = \log\left( \dfrac{1-p}{p} \right) \quad \text{and}\quad   w_t = \log\left( \dfrac{1-q}{q} \right),
\end{equation}
since the rate $p$ of coherent errors can differ from the rate $q$ of readout errors. To ensure that the correction operator brings the state back to the code space, the last measurement cycle is assumed to be free of any readout errors, and the decoder is aware of this assumption.

The logical error rate is obtained by averaging over all (noisy) syndromes and the $\boldsymbol{\theta}$ angle distribution as well,
\begin{equation}\label{eq:pL-surface-code}
    p_L (\sigma) = \sum_{\boldsymbol{s},\boldsymbol{s^\prime}} 
P\left(\boldsymbol{s}\right) P\left(\boldsymbol{s}\to\boldsymbol{s^\prime}\right) \Big\langle \sin^2\theta_L (\boldsymbol{s},\boldsymbol{s^\prime},\boldsymbol{\theta}) \Big\rangle_{\boldsymbol{\theta}} .
\end{equation}

\subsection{Numerical results} \label{subsec-5D}

Standard Pauli errors in quantum error correction can be simulated efficiently with Clifford simulators \cite{gottesman1997stabilizer,Aaronson_2004}. Unfortunately, to simulate quasistatic coherent errors we could not use this tool. To explore the effect of coherent errors in the surface code, several numerical methods have been considered. Brute-force simulations of small systems \cite{Tomita_2014,O_Brien_2017}, tensor-network simulations for slightly larger systems \cite{Darmawan_2017}, and Fermionic Linear Optics (FLO) methods \cite{Bravyi_2018,VennPRR2020,Marton2023coherenterrors} have emerged as possible options. In this work, we used the FLO method, as implemented in \cite{Marton2023coherenterrors}, to simulate the surface code under quasistatic phase damping and readout errors for code sizes up to $d = 17$. For further details of the simulation method, we recommend the appendices of Refs.~\cite{Bravyi_2018,Marton2023coherenterrors}.

We sampled the distribution of the logical rotation angle, and computed the logical error rate in the following way:
\begin{equation}
    p_L (\sigma) \approx \dfrac{1}{N_\text{sample}}  \sum_{i=1}^{N_\text{sample}} \sin^2\left(\theta_L (\boldsymbol{s}^{(i)}, \boldsymbol{s^\prime}^{(i)},\boldsymbol{\theta}^{(i)})\right) ,
\end{equation}
where $N_\text{sample}$ denotes the sample size.

\begin{figure}
    \centering
    \includegraphics[width=0.46\textwidth]{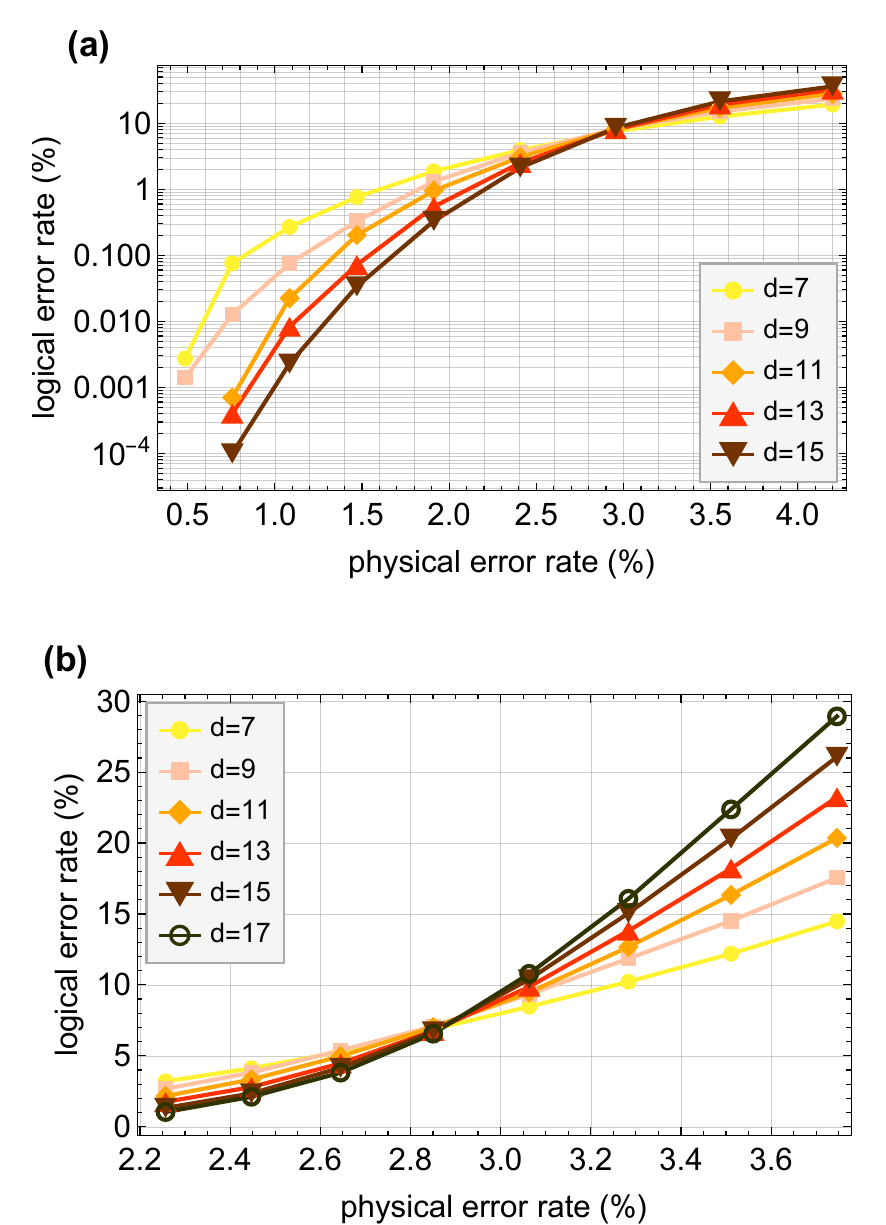}
    \caption{Surface code threshold with quasistatic phase damping and readout errors for $p=q$. Data points show the numerically obtained logical error rate (maximum infidelity), for different code distances $d$, with $p = q = \langle\sin^2\theta\rangle$ being the physical error rate $p$ and readout error rate $q$, that are kept equal for this plot. Each point is a result of $N_\text{sample} = 10,000\times d \times 100$ simulations: we sampled the multi-cycle syndrome $\boldsymbol{s}$ for $10,000$ random $\boldsymbol{\theta}$ angles, simulating $100$ random readout error configurations in each case. (a) Logical error rate for a wide range of physical error rates. (b) Zoom-in of (a) into a narrow window of physical error rates in the vicinity of the threshold.
    The graph confirms the existence of a threshold at $p_\text{th} \approx 2.85\%$.
}
\label{fig:surface-code-threshold}
\end{figure}

Fig.~\ref{fig:surface-code-threshold}a shows that by increasing the code distance, we observe threshold behavior. When the rates of physical and readout errors are the same, $p = q = \langle \sin^2\theta \rangle$, we find that the threshold is $p_{\text{th}} \approx 2.85\%$ (see Fig.~\ref{fig:surface-code-threshold}b). 
This is close to the threshold with Pauli Z errors and readout errors, which for the toric code is $2.93\pm 0.02\%$  \cite{Wang_2003}. 
We also carried out Monte Carlo simulations to investigate phase-flip errors. Interestingly, we have found that the threshold value coincides with the above mentioned $2.85\%$, however, the corresponding logical error rate is different for the two error models: in our case $p_L(p_{\text{th}})\approx 7\%$, but for the combination of phase-flip and readout errors it is slightly higher, around $8.5\%$.
This relative difference is sustained throughout the physical error rate window of Fig.~\ref{fig:surface-code-threshold}b (not shown).

\begin{figure}
    \includegraphics[width=0.45\textwidth]{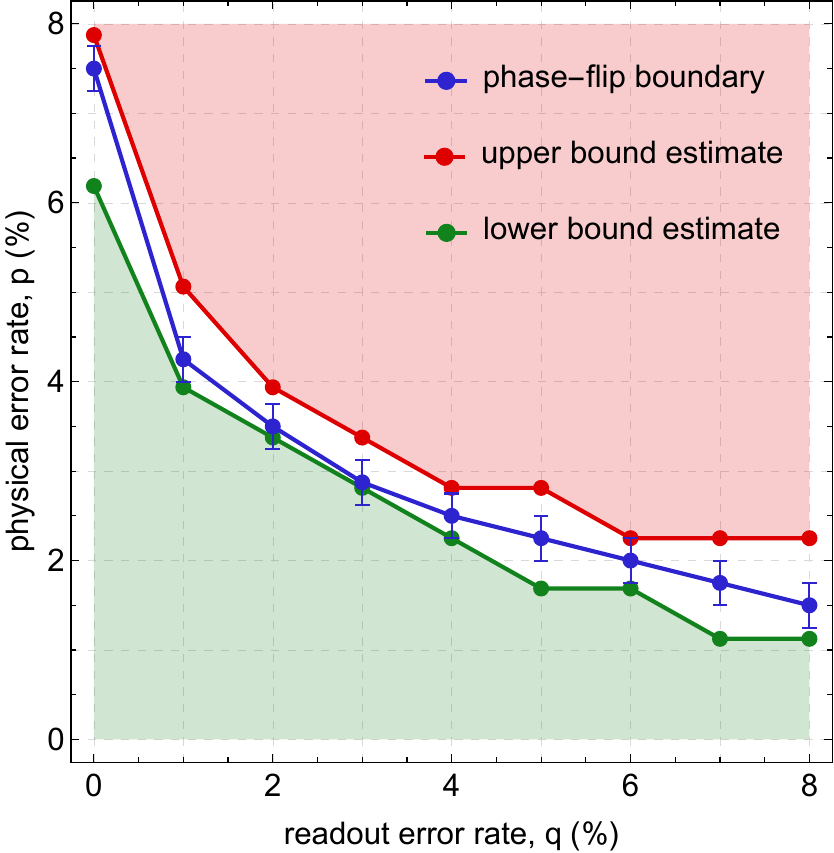}
    \caption{Surface code threshold with quasistatic phase damping and readout errors. On the $(p,q)$ plane, threshold lines were identified by simulating 16 different values of $p$ (equidistant in $0<p\leq 9\%$) for a fixed $q$ value, for code distances $d=7,9,11,13,15$. 
    For each $q$ and $d$, a lower and an upper bound for the $p$ threshold was estimated.
    The lower estimate is the greatest value of $p$ at which the maximum infidelity decreases as the code distance increases. The upper estimate is the smallest value of $p$ at which the logical error rate increases as the code distance increases. We also carried out Monte Carlo simulations to investigate phase-flip errors and numerically determined the threshold values for asymmetric $p$ and $q$ values.}
\label{fig:pq-plane}
\end{figure}

We have also investigated how varying the rate $p$ of random coherent errors and $q$ of readout errors independently affect the threshold of the surface code. Results are shown in \ref{fig:pq-plane}. For many pairs of $p$ and $q$ we numerically assessed whether scaling up the surface code decreases the logical error rate (scalable quantum error correction, green area) or increases it (unscalable quantum error correction, red area). The threshold should be in between these regions. 
For example, if the physical error rate is at the $1\%$ level, the surface code is quite robust against readout errors, scalable even with relatively high $q\approx 8 \%$. However, if the readout error rate is at the $1\%$ level, the surface code requires the coherent error rate to be below $4 \%$.

\section{Results for the distance-3 surface code, and their relation to experiments} \label{sec-6}

In this section, we discuss the implications of our results for the smallest surface code ($d=3$) capable of error correction and their relation to experiments.

For physical error rates much below the threshold, $p \ll p_{\text{th}} \approx 2.85\%$, numerical simulations could only provide accurate results by significantly increasing the number of samples which is not feasible. However, it would also be interesting to see how the surface code performs in this regime.
For the distance-3 surface code, we could carry out analytic calculations using a MWPM-based look-up table decoder, setting the weights of 'spacelike' and 'timelike' edges equal. For $3$ rounds of error correction with readout error rate equal to physical error rate, $q=p$, we find that the logical error rate up to leading order is given by
\begin{equation}
    p_L^{(d=3)} \approx 118 p^2.
\end{equation}
Interestingly, this expression is the same for independent phase-flip and readout errors. Thus, for small error rates, we expect similar behavior for the two models.

\begin{figure}
    \includegraphics[width=0.45\textwidth]{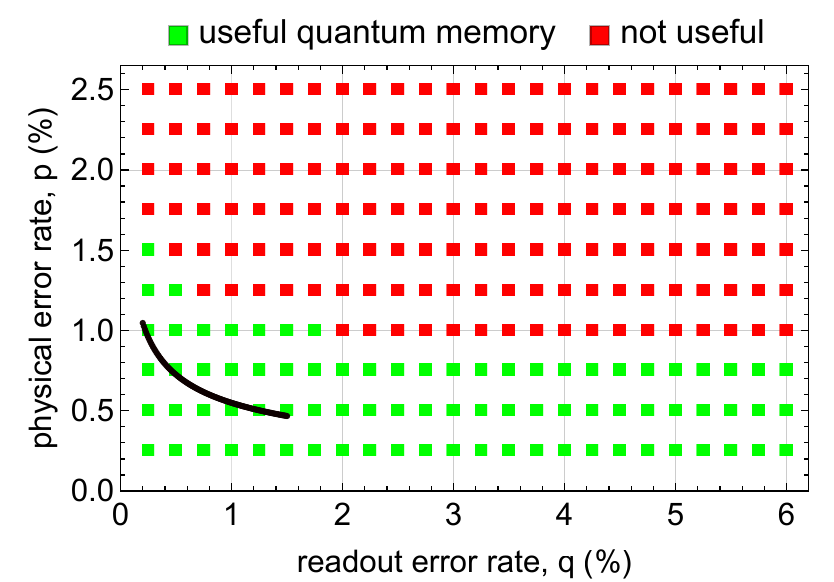}
     \caption{Comparison of logical and physical error rates on the $(p,q)$ plane for the surface code with code distance $d=3$. The green color indicates $p_L < p$, while the red color is for $p_L \ge p$. The black line shows estimated physical and readout error rates based on state-of-the-art spin qubit experiment, Ref.~\cite{Takeda_2024}. It corresponds to readout time $0.5-0.7\, \mu$s.
     }
\label{fig:surface-code-d-3-break-even}
\end{figure}

We consider an error correction experiment useful when the logical error rate of the encoded qubit is lower than the error rate of a single physical qubit, $p_L < p$. 
To study when this break-even condition is achieved, we compute the logical error rate $p_L$ for several fixed values of physical error rate $p$ and the readout error rate $q$ for $d=3$. Results are shown in Fig.~\ref{fig:surface-code-d-3-break-even}. Again, we see that coherent errors are more harmful to the surface code quantum memory than readout errors: to achieve break-even, the physical error rate needs to be pushed below $1 \%$, however, in this regime, readout errors up to $6\%$ are tolerated. 

We now connect these results to experimentally relevant error parameters, focusing on semiconductor spin qubits, to estimate the expected performance of a small surface code experiment based on currently available spin qubit hardware.
Our considerations expand upon earlier theoretical quantum error correction studies focusing on spin qubits \cite{Nigg_2017,Helsen_2018,Buonacorsi_2019,Cai_2019,Rispler_2020,Rozgonyi_2023,hetényi2023tailoring}.
We remark that the first error correction experiments with semiconductor spin qubits realized the phase-flip repetition code \cite{Takeda_2022,Riggelen_2022} with a single round of measurement and without feedforward; however, the surface code has yet to be realized.

The solid black curve in Fig.~\ref{fig:surface-code-d-3-break-even} shows our estimate for the physical and readout error rates, based on recent experimental data \cite{Takeda_2024} obtained for spin qubit readout in a double quantum dot. 
Points of the curve correspond to different measurement integration times, as we explain below. 
The black curve lies inside the break-even region (green), indicating that
scaling up the hardware with the same high-quality components could make useful surface-code-based quantum error correction attainable with semiconductor spin qubits.

The black curve in Fig.~\ref{fig:surface-code-d-3-break-even} is obtained from the data \cite{Takeda_2024} as follows.
We connect our coherent error strength $\sigma$, and the corresponding physical error rate $p$, to the experimentally available inhomogeneous dephasing time $T_2^*$ and measurement integration time $T_\text{meas}$ by assuming that the Larmor angular frequency $\omega_L$ of each idling data qubit exhibits spatially uncorrelated, Gaussian, quasistatic fluctuations, with standard deviation $\sigma_{\omega_L} = \sqrt{2}/ T_2^*$. This detuning of the Larmor frequency of the $j$th data qubit of the code implies a small random rotation $\theta_j$ around the z axis on that qubit. Typically, the duration of a quantum error correction cycle is dominated by the measurement integration time $T_{\text{meas}}$ of the measure qubits, therefore we estimate the standard deviation $\sigma$ of the angle $\theta_j$ of the coherent error as  
\begin{equation} \label{eq:delta-phi}
    \sigma(T_\text{meas}) = \sigma_{\omega_L}\cdot T_{\text{meas}} = \sqrt{2}\dfrac{T_{\text{meas}}}{T_2^*}.
\end{equation}
From this result, we express the physical error rate $p(T_\text{meas})$ using Eq.~(\ref{eq:rep-2-single-cycle-p}). 

We can also provide a realistic estimate for the dependence of the readout error rate $q$ on the measurement integration time. 
For this, we use experimental data, namely, the data set corresponding to the initial state $T_-$ in Fig.~3d of Ref.~\cite{Takeda_2024}.
In the measurement integration time range $[0.5,0.7]\, \mu$s, the data is well approximated by this power law: 
\begin{equation}
    q(T_\text{meas}) \approx  
    \left(
    \frac{\tau}{T_{\text{meas}}}
    \right)^5,
\end{equation}
where $\tau = 0.21\,\unit{\mu s}$ is a time constant.

The solid black curve in Fig.~\ref{fig:surface-code-d-3-break-even} shows the parametric dependence of $q$ and $p$ on $T_\text{meas}$, based on the considerations and numerical factors of the previous two paragraphs, and inhomogeneous dephasing time value \cite{Takeda_2024} of $T_2^\ast = 10\, \unit{\mu s}$.
The black curve lies in the green region, where the logical error rate is lower than the physical error rate.
Based on these considerations, we conclude that although it is challenging to perform a 9-qubit surface code experiment, semiconductor spin qubits have the potential to achieve break-even in the near future.
In experiments, where the readout is significantly slower than in the one we cited, it could be fruitful to convert high readout fidelity to shorter readout time duration \cite{Rispler_2020}, since the surface code is more robust against readout errors than coherent errors.
Note that the above considerations can be significantly improved with the goal of a more realistic description of quantum memory experiments, e.g., by taking into account gate errors both on data and measure qubits, etc.

Finally, we also relate our simulation results to superconducting quantum computing platforms. On the one hand, transmons are designed to be protected from dephasing \cite{Koch}, and hence their decoherence is often dominated by relaxation, which is excluded from our model. On the other hand, alternative superconducting qubits have been developed  \cite{Fedorov,Earnest,Lin,GyenisExperiment}, in which qubit relaxation is suppressed and pure dephasing due to charge noise or flux noise is the dominant decoherence mechanism. 
(For an informative summary, see Ref.~\cite{GyenisPRXQuantum}, and Table 1 therein.)
For such superconducting qubits, the quasistatic phase damping model is a good starting point.
Therefore, it is an interesting and relevant future task to assess the performance of quantum error correction based on such qubits, utilizing the efficient simulation methodology (fermionic linear optics) applied in this paper. 

\section{Conclusions} \label{discussion}

In the context of multi-cycle quantum error correction, we introduced the error model of quasistatic phase damping, which is a simplified model describing Larmor-frequency fluctuations due to $1/f$ noise. 
These Larmor-frequency fluctuations amount to unwanted coherent rotations whose axis is uniform, but the rotation angle is random across the qubit register. 
We showed that this quasistatic phase damping error is equivalent to independent single-qubit Pauli phase-flip errors in the case of a single cycle of error detection or error correction.
However, for multiple cycles, quasistatic phase damping is distinct from a phase-flip error.

Furthermore, we numerically investigated the performance of the surface code as quantum memory in the presence of quasistatic random coherent errors on the physical qubits as well as (phenomenological) readout errors. We utilized the Fermionic Linear Optics simulation framework to perform large-scale numerical experiments and established the surface code error correction threshold. The threshold is found to be close to that with independent phase-flip errors and readout errors. However, at the threshold, the logical error rate for quasistatic phase damping combined with readout errors is lower than for independent phase flips and readout errors.
This suggests that in terms of surface code quantum memory, quasistatic phase damping is less harmful.
Threshold line estimates in the plane spanned by the strengths of the coherent error and the readout error also show that quasistatic phase damping and the independent phase-flip error model have similar threshold values.

We have evaluated the effects of our error model on the distance-3 surface code, and established the break-even boundary line for this code on the parameter plane of the physical error rate and the readout error rate. 
Finally, we have discussed the relevance of our results to solid-state qubit platforms, including semiconductor spin qubits and superconducting qubits. 
Our results are expected to provide useful design guidelines for future experiments based on these quantum computing platforms.

We note that a possible extension of our error model is the combination of quasistatic phase damping and random Pauli errors. Combining random phase-flip errors with quasistatic phase damping, as considered in Ref.~\cite{Huang_2019}, can be an interesting future direction. Also including random Pauli $X$ and $Y$ errors, as, e.g., in depolarizing noise, however, is unfortunately beyond the scope of our approach,
due to the constraints of the FLO simulation.

\section*{Acknowledgments}

We thank A.~Gyenis and B.~Het\'enyi for helpful discussions. 
This research was supported by the Ministry of Culture and Innovation and the National Research, Development and Innovation Office within the Quantum Information National Laboratory of Hungary (Grant No. 2022-2.1.1-NL-2022-00004). 
This research has been supported by the Horizon Europe research and innovation programme of the European Union through the IGNITE project and by the HORIZON-CL4-2022-QUANTUM01-SGA project 101113946 OpenSuperQPlus100 of the EU Flagship on Quantum Technologies.
This research was supported by the National Research, Development and Innovation Office via the OTKA Grant No. 132146, and the ”Frontline” Research Excellence Programme 
(Grant No. KKP133827).
This project has received funding from the HUN-REN Hungarian Research Network.

\appendix

\section{Calculation details for the 2-qubit repetition code for two cycles of error detection}

Here, we list the 16 cycle-resolved error scenarios and the corresponding merged error operators for the 2-qubit repetition code, for two cycles of error detection. For comparison, we provide the probabilities obtained from a phase-flip error model alongside the coherent error amplitudes for each error scenario, see Table~\ref{tab:rep-2-errors}.

\begingroup
\setlength{\tabcolsep}{8pt}
\begin{table*}
\centering
\begin{tabular}{ccccccc}
\hline\hline
No. & $\boldsymbol{E}$ & $\hat{E}(\boldsymbol{E})$ & $\boldsymbol{s}$ & $P(\boldsymbol{E})$ & $\Tilde{P}(\boldsymbol{E})$ \\ \hline 
1  &  $(\hat{\mathds{1}},\hat{\mathds{1}})$       &  $\hat{\mathds{1}}$    &  $(+1,+1)$ & $(1-p)^4$      & $1-4 p+11 p^2-26 p^3+46 p^4-60 p^5+56 p^6-32 p^7+8 p^8$  \\ 
2  &  $(\hat{Z}_A \hat{Z}_B,\hat{Z}_A \hat{Z}_B)$ &  $\hat{\mathds{1}}$    &  $(+1,+1)$ & $p^4$          & $p^2-6 p^3+26 p^4-52 p^5+56 p^6-32 p^7+8 p^8$            \\ 
3  &  $(\hat{\mathds{1}},\hat{Z}_A \hat{Z}_B)$    &  $\hat{Z}_A\hat{Z}_B$  &  $(+1,+1)$ & $(1-p)^2 p^2$  & $2 p^2-12 p^3+34 p^4-56 p^5+56 p^6-32 p^7+8 p^8$         \\ 
4  &  $(\hat{Z}_A \hat{Z}_B,\hat{\mathds{1}})$    &  $\hat{Z}_A\hat{Z}_B$  &  $(+1,+1)$ & $(1-p)^2 p^2$  & $2 p^2-12 p^3+34 p^4-56 p^5+56 p^6-32 p^7+8 p^8$         \\ 
5  &  $(\hat{Z}_A,\hat{Z}_A)$                     &  $\hat{\mathds{1}}$    &  $(-1,+1)$ & $(1-p)^2 p^2$  & $4 p^2-16 p^3+36 p^4-56 p^5+56 p^6-32 p^7+8 p^8$          \\ 
6  &  $(\hat{Z}_B,\hat{Z}_B)$                     &  $\hat{\mathds{1}}$    &  $(-1,+1)$ & $(1-p)^2 p^2$  & $4 p^2-16 p^3+36 p^4-56 p^5+56 p^6-32 p^7+8 p^8$          \\ 
7  &  $(\hat{Z}_A,\hat{Z}_B)$                     &  $\hat{Z}_A\hat{Z}_B$  &  $(-1,+1)$ & $(1-p)^2 p^2$  & $2 p^2-12 p^3+34 p^4-56 p^5+56 p^6-32 p^7+8 p^8$           \\ 
8  &  $(\hat{Z}_B,\hat{Z}_A)$                     &  $\hat{Z}_A\hat{Z}_B$  &  $(-1,+1)$ & $(1-p)^2 p^2$  & $2 p^2-12 p^3+34 p^4-56 p^5+56 p^6-32 p^7+8 p^8$           \\ 
9  &  $(\hat{\mathds{1}},\hat{Z}_A)$              &  $\hat{Z}_A$           &  $(+1,-1)$ & $(1-p)^3 p$    & $p-6 p^2+19 p^3-40 p^4+58 p^5-56 p^6+32 p^7-8 p^8$        \\ 
10 &  $(\hat{Z}_A \hat{Z}_B,\hat{Z}_B)$           &  $\hat{Z}_A$           &  $(+1,-1)$ & $(1-p) p^3$    & $-p^2+9 p^3-30 p^4+54 p^5-56 p^6+32 p^7-8 p^8$             \\ 
11 &  $(\hat{\mathds{1}},\hat{Z}_B)$              &  $\hat{Z}_B$           &  $(+1,-1)$ & $(1-p)^3 p$    & $p-6 p^2+19 p^3-40 p^4+58 p^5-56 p^6+32 p^7-8 p^8$        \\ 
12 &  $(\hat{Z}_A \hat{Z}_B,\hat{Z}_A)$           &  $\hat{Z}_B$           &  $(+1,-1)$ & $(1-p) p^3$    & $-p^2+9 p^3-30 p^4+54 p^5-56 p^6+32 p^7-8 p^8$            \\ 
13 &  $(\hat{Z}_A,\hat{\mathds{1}})$              &  $\hat{Z}_A$           &  $(-1,-1)$ & $(1-p)^3 p$    & $p-6 p^2+19 p^3-40 p^4+58 p^5-56 p^6+32 p^7-8 p^8$        \\ 
14 &  $(\hat{Z}_B,\hat{Z}_A\hat{Z}_B)$            &  $\hat{Z}_A$           &  $(-1,-1)$ & $(1-p) p^3$    & $-p^2+9 p^3-30 p^4+54 p^5-56 p^6+32 p^7-8 p^8$            \\ 
15 &  $(\hat{Z}_A,\hat{Z}_A\hat{Z}_B)$            &  $\hat{Z}_B$           &  $(-1,-1)$ & $(1-p) p^3$    & $-p^2+9 p^3-30 p^4+54 p^5-56 p^6+32 p^7-8 p^8$             \\ 
16 &  $(\hat{Z}_B,\hat{\mathds{1}})$              &  $\hat{Z}_B$           &  $(-1,-1)$ & $(1-p)^3 p$    & $p-6 p^2+19 p^3-40 p^4+58 p^5-56 p^6+32 p^7-8 p^8$        \\ \hline\hline
\end{tabular}
\caption{Summary of cycle-resolved error scenarios $\boldsymbol{E}$, merged error operators $\hat{E}(\boldsymbol{E})$, syndromes $\boldsymbol{s}$, independent Pauli phase-flip probabilities $P(\boldsymbol{E})$, and coherent error amplitudes $\Tilde{P}(\boldsymbol{E})$ (written as a function of $p = \langle\sin^2\theta\rangle$) calculated from Eq.~\eqref{eq:tilde-P-E}, for the 2-qubit repetition code, for two cycles of error detection.}
\label{tab:rep-2-errors}
\end{table*}
\endgroup

\clearpage
\bibliography{QEC}

\end{document}